\documentclass[floatfix,preprintnumbers,superscriptaddress,showpacs,
               prl,preprint]{revtex4}

\usepackage{graphicx}
\usepackage{dcolumn}
\usepackage{amsmath, amsfonts, amssymb, bm}

\begin{document}

\bibliographystyle{apsrev} 
\preprint{CSIC/08/ICMM/Torroja/Fe-H-jj}

\title{Hydrogen in $\alpha$-iron: stress and diffusion} 

\author{J. Sanchez}
\author{J. Fullea}
\author{C. Andrade}
\affiliation{
Instituto de Ciencias de la Construccion, "Eduardo Torroja" (CSIC) 
c/ Serrano Galvache 4, E-28033 Madrid (SPAIN)
}

\author{P.L. de Andres}
\affiliation{
Instituto de Ciencia de Materiales de Madrid (CSIC)
E-28049 Cantoblanco, Madrid (SPAIN)
}

\date{\today}

\begin{abstract}
First-principles density-functional theory has been used to investigate 
equilibrium geometries, total energies, and diffusion barriers for H 
as an interstitial impurity absorbed in $\alpha$-Fe.
Internal strains/stresses upon hydrogen absorption are a crucial factor 
to understand preferred absorption sites and diffusion. 
For high concentrations, H absorbs near the octahedral site favoring a
large tetragonal distortion of the BCC lattice. For low concentration, H 
absorbs near the tetrahedral site minimizing the elastic energy
stored on nearby cells.
Diffusion paths depend on the concentration regime too; hydrogen
diffuses about ten times faster in the distorted BCT lattice. 
External stresses of several GPa modify barriers
by $\approx$ 10\%, and diffusion rates by $\approx$ 30\%.
\end{abstract}

%


\maketitle

\section{\label{sec:intro} Introduction and Methodology}
\begin{figure}
\includegraphics[clip,width=0.8\columnwidth]{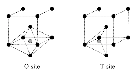}
\caption{
High symmetry sites for H absorption inside $\alpha$-iron:
the octahedral (O-site) and tetrahedral (T-site) sites (grey) are shown
along with iron atoms in the BCC lattice (black).
}
\label{FigTO}
\end{figure}

Hydrogen embrittlement is believed to be one of the main reasons for
cracking of steel structures under stress\cite{eliaz02,elices04,liang03}. 
High strength steels often include a ferritic core made of $\alpha$-iron
(body centered cubic lattice, BCC). To control and prevent the
cracking of steel it is necessary to understand 
the chemical and physical properties of hydrogen inside BCC iron.
In particular, it has been argued in the literature that the
size of the interstitial H (e.g. $\approx 0.3-0.4$ {\AA}\cite{pauling})
is too large to fit and diffuse easily in the iron BCC
lattice\cite{barret}. This argument has also been used to favor
absorption on the T-site because it is larger than the 
O-site\cite{griessen88}. 
Arguments based in 'apparent' sizes of atoms and
a static lattice yield an useful basic approximation but
leave out important factors.
Only a truly microscopic understanding of hydrogen dissolved in iron, 
including diffusion processes and fully accounting for forces and
stresses inside the host matrix 
could lead to answers to the many questions arising from experiments.
Internal stresses are therefore key to understand 
the preferred absorption site. 
Furthermore, external stresses can be
used to modify diffusion barriers
because can help to stabilize/de-stabilize strains originated inside 
the material to accommodate the absorbed impurity.
Owing to its interest, hydrogen absorbed in iron has been studied
from different approaches, both from a theoretical and experimental
point of view.
Unfortunately, a clear consensus about fundamental questions, like the 
nature of the equilibrium absorption site, the reasons for H to prefer
some regions over others, or how to modify the diffusion barriers
has not been reached yet. 
Many theoretical papers have favored hydrogen
absorbed in the tetrahedral site
(T-site)\cite{norskov82,liu85,demangeat87,louie98,hoffman99,carter04}, 
some have preferred the octahedral one 
(O-site)\cite{zheng89}, others have reported that they are almost 
equivalent\cite{nieminen84} (Fig. ~\ref{FigTO}). 
From an experimental point of view, it has been generally believed
that hydrogen has low solubility and high mobility in bcc-iron.
However, some recent experiments have attained high
concentrations\cite{antonov98,castellote07}, at least locally.
Furthermore, experimental evidence has been reported showing
that hydrogen might at least partly be found on the O-site\cite{antonov98}.
Finally, diffusion barriers that change with the amount of H admitted into 
the material have been reported in the literature\cite{kiuchi83}.
In our calculations, we find that both sites are indeed quite similar 
in energies; 
the delicate energetic balance between them depending on the stress 
distribution created by the interstitials. In turn, this depends on 
the hydrogen concentration, that can be related to locally inhomogeneous
regions or more generally to its homogeneous averaged value.

Our methodological approach simulates the atomic absorption and 
diffusion of hydrogen in the body centered cubic iron lattice from 
first-principles. Absorption in high-symmetry sites for different 
hydrogen loads (densities) has been explored and the internal 
strains/stresses have been calculated.  
Diffusion barriers between these stationary absorption sites have been 
obtained, and the effect of external stresses on diffusion coefficients 
has been analyzed. 
Our {\it ab-initio} calculations are based on the Density Functional Formalism 
(DFT)\cite{hohenberg65} and the theory of pseudopotentials\cite{vanderbilt90}. 
A periodic supercell is considered to minimize absorbate-absorbate
interactions (periodicity is $n \times n \times n$ with respect to a single 
cubic BCC unit cell -UC-). 
One H atom has been located in this supercell, determining the ratio 
H:Fe 
from $\frac{1}{2}$ for the $1\times 1\times 1$ to $\frac{1}{54}$ for 
the $3\times 3\times 3$.
Within this theoretical framework the accuracy of calculations
mainly depend on the quality of the pseudopotential,
the energy cutoff ($E_{ctf}$), and the density
of the k-points mesh in the irreducible part of the Brillouin zone
(a Monkhorst-Pack set of special k-points\cite{monk}). 
Calculations are performed with CASTEP\cite{payne02,accelrys}, 
where a plane-wave basis is used to expand electronic states 
(spin-polarized bands are considered to account for magnetism in iron).
An ultrasoft pseudo-potential (3d$^{6}$ 4s$^{2}$) 
including a nonlinear core correction 
is used to describe iron atoms. 
The exchange and correlation contribution to the total energy is computed 
within a generalized gradients approximation given by Perdew, Burke and
Ernzerhof\cite{pbe}. 
An all-bands variational method is used to achieve 
self-consistency\cite{edft} (
to improve convergence
a smearing width of $0.1$ eV has been used). 
Minimum thresholds to consider the calculations converged after optimization 
of the geometry are: 
variations in the total energy $\le 10^{-5}$ eV, 
maximum residual force on any atom $\le 0.01$ eV, 
maximum residual stress on the UC $\le 0.02$ GPa, 
maximum change in any atom position $\le 0.001$ {\AA}, 
and maximum change in any  cell vector $\le 0.01$ {\AA}. 
These thresholds are enough for our purposes, 
but on some particular cases lower thresholds were considered 
when higher precision was desirable. 

\begin{figure}
\includegraphics[clip,width=0.8\columnwidth]{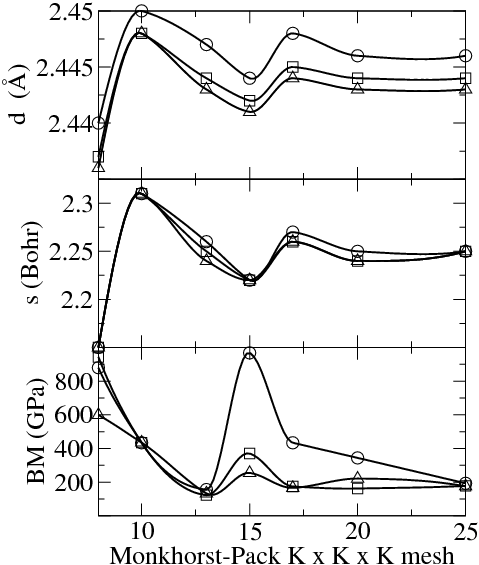}
\caption{Convergence of primitive unit cell parameter ( d in {\AA})
-upper panel-,
magnetization (in units of the Bohr magneton) -middle panel-, 
and  bulk modulus (GPa) -lower panel- versus
the energy cut-off (circles:
330 eV, squares: 350 eV, triangles: 370 eV), and 
the $K \times K \times K$ index for the Monkhorst-Pack mesh
($K=25$ results in  a mesh of 455 points sampling  
the Brillouin zone).
}
\label{FigCONV}
\end{figure}

\section{Reference calculations for $\alpha$-iron and H}
The adequacy of the theoretical framework and the different convergence 
thresholds has been tested by computing a number of known properties for 
clean $\alpha$-iron and atomic and molecular hydrogen separately. 
For iron, a global energy minimum is found for a BCC lattice
(hereafter referred as the $1 \times 1 \times 1$) with 
$a=b=c=2.815$ {\AA} ($\alpha=\beta=\gamma=90^{o}$),
a predicted magnetization of 2.25 $\mu_{B}$,
a bulk modulus of 175 GPa, 
and a Poisson ratio of 0.30.
Residual forces on the atoms in the BCC unit
cell are $\le 4 \times 10^{-6}$ eV/{\AA}, 
residual stress is $\le 9 \times 10^{-4}$ GPa,
and the total energy has been converged to $\le 10^{-6}$ eV).  
The computed magnitudes are to be compared with the experimental
values 2.867 {\AA}, 2.22 $\mu_{B}$, 170 GPa, and 
0.29\cite{handbook}.
Fractional errors are 1.8\%, 1.4\%, 3\%, and 4\% respectively,
showing the accuracy of this kind of theoretical modeling to 
describe relevant properties of BCC iron.
It is interesting, however, to notice that these quantities show oscillatory 
convergence as a function of the energy cutoff ($E_{ctf}$) and the number
of k-points;
convergence of physical properties has to be established 
carefully to make sure the oscillations are damped enough
as to avoid agreements that might not be stable 
(Fig. ~\ref{FigCONV}). 
Properly converged values were obtained with 
$E_{ctf}=375$ eV and a $25 \times 25 \times 25$ Monkhorst-Pack (MP)
mesh (455 points in the irreducible part of the Brillouin zone).
However, values derived with 
$E_{ctf}=375$ eV and a $14 \times 14 \times 14$ mesh 
can be considered accurate enough for our purposes and most
of our calculations below for the $1 \times 1 \times 1$ BCC unit cell
have been performed with these parameters.
Calculations in $n \times n \times n$ unit cells (n=2, 3)
have been performed with $4 \times 4 \times 4$ and
$1 \times 1 \times 1$ MP meshes respectively. 
These theoretical calculations yield a detailed microscopic
picture for the chemistry and the physics governing the material:
iron atoms form metallic bonds of length 2.44 {\AA},
favoring a ferromagnetic alignment of spins mostly residing in the regions
around the atoms.
Bonding is such that the atomic s level of iron is depopulated in favor of
d and p-levels (a Mulliken analysis shows that $\approx 1.3$ e is transferred).
Although this work is not concerned with face-centre cubic iron
($\gamma$-iron), we notice that
this formalism predicts that the next stable phase of iron (FCC)
is about $0.1$ eV/atom higher in energy, with
$a=b=c=3.479$ {\AA} (experimental value is $3.647$ {\AA},
a 5\% discrepancy) , and a magnetization lower by about a factor of 
$2$ due to disorder in the spin orientations. 
Calculations for H and H$_{2}$ have been performed on a 
$10 \times 10 \times 10$ {\AA} super-cell (only
the $\Gamma$ point in the BZ),  
predicting a bond length and a binding energy for the molecule 
within 0.5\% and 5\% from experimental values. 
The interaction between isolated atoms of H and Fe is moderately weak, 
with a bonding distance of 1.54 {\AA} and a binding energy of 
$E_{B} = E_{FeH} - (E_{Fe}+E_{H}) = -2.41 eV$
(notice that the binding energy is quoted from
the electronic energy alone and 
zero point vibrations have not been considered).
This is corroborated by the very weak interaction of H with Fe$_{2}$; 
the binding
energy stays nearly the same ($-2.43$ eV) in a geometrical configuration 
where H prefers to interact with only one of the Fe in the dimer.
Molecular H$_{2}$ has been reported to dissociate 
on iron surfaces\cite{carter04}, in addition
we do not find tendency for hydrogen 
atoms inside iron to re-form molecules. As an example, we compare for the
$2 \times 2 \times 2$ unit cell (Fe$_{16}$H$_{2}$) the total
energies of two H atoms located in nearby two different octahedral sites
(H-H distance is $\approx 2$ {\AA})
and a single H$_{2}$ molecule. The dissociated state is favored by 
$E_{2H}-E_{H_{2}}=-3.3$ eV (we notice that the energy gain comes partially 
from the smaller stress that the two separated H atoms introduce
in the iron lattice 
compared with the one originated from the single H$_{2}$ molecule).
Therefore, we conclude that it is not necessary to study
processes related to molecular H$_{2}$ inside $\alpha$-iron.

\section{Hydrogen in iron: high concentration}
Even if the density of hydrogen absorbed in iron is expected to be small 
on the average, it can accumulate locally on some regions where the density 
can be high. 
Indeed, important changes in the elastic properties of iron are found in 
these regions.
Furthermore, experimental evidence of high density phases
have been reported in the literature\cite{antonov98,castellote07}.
Therefore, we find it interesting to analyze the stoichiometry Fe$_{2}$H.
In the BCC lattice two high-symmetry sites (O-site and T-site) compete to locate
the interstitial atom, and diffusion through
the iron matrix takes place along paths joining them.
We simultaneously optimize all the geometrical parameters
to reach a global minimum in the total energy and small
residual forces in any atom in the model.
The three vectors defining the periodic lattice have been kept
orthogonal ($\alpha=\beta=\gamma=90^{o}$). The lengths of these 
vectors have been optimized taking into account the following cases:
(a) independent optimization
(resulting in tetragonal systems), (b) optimization subject
to the condition $a=b=c$ (cubic system), and (c) no optimization,
merely scaling the clean BCC unit cell (the system is under stress,
specially for high concentrations of the interstitial). 
Starting with model (a) and allowing the lattice to relax
completely, we have reached a best configuration with
residual forces 
$\le 10^{-4}$ eV/{\AA} and residual stresses
$\le 5 \times 10^{-4}$ GPa
(the energy change was $\le 2 \times 10^{-6}$ eV and the
maximum final displacement of any atom
was $\le 5 \times 10^{-4}$ {\AA}).
Under these conditions, the O-site is the
global minimum, the T-site being higher in energy by $53$ meV
(Table~\ref{TabDFT}). 
To reach this configuration, the lattice suffers a large
body centered tetragonal (BCT) distortion, keeping
the P4/MMM symmetry group (16 elements), while
the optimum geometry for the fully relaxed 
lattice for H in the T-site increases the symmetry from
C2 (4 elements) to P-4M2 (8 elements).
Our preferred absorption site is different from 
the one recently obtained by Jiang 
and Carter for the same density using a similar formalism\cite{carter04}. 
There are several possible sources for this discrepancy, e.g. the 
pseudopotentials, the exchange and correlation functional, or even the 
fine details of how different codes reach a self-consistent solution 
and compute forces on the atoms.  We believe these are all unlikely 
sources to explain the different final configuration due to the excellent 
performance when comparing different physical magnitudes with experimental 
data or just between different theoretical methods of similar quality.
Jiang and Carter predict for Fe$_{2}$H that the T-site is preferred over 
the O-one by $10$ meV, while we have found the reversed conclusion by 
$53$ meV. The overall $63$ meV discrepancy is indeed a small one, 
but it seems to be above the noise level expected from the convergence
tests. The reason for the difference is most likely related
to the final optimum configuration obtained in each calculation
(our final configuration is given in Table ~\ref{TabGEOM1}).
The energy landscape is a complex one, and the algorithm searching for
a global minimum could be trapped in nearby local ones.
We have made an effort to minimize residual forces and stresses as much as 
possible. To search for a global minimum we have used the 
Broyden-Fletcher-Goldfarb-Shanno algorithm that is known to be 
a robust one and has been successfully applied both in electronic
structure calculations and in Low-Energy Electron Diffraction
optimization of geometries\cite{recipes,bfgs,maria06}. 
The resulting BCT distortion for H in the O-site 
is indeed quite big and might be hidden in a 
difficult landscape, we notice that our optimization under the restriction 
of an isotropic deformation of the unit cell ($a=b=c$) results in the T-site 
as the preferred one by $452$ meV. Finally, we should comment that although 
the BCT distortion seems significant, in particular because it would favor
the BCC to FCC transformation, the discussion here is based solely on 
electronic energies, and other factors might play an important role too.
Configurational disorder is expected to favor T-sites.
The relative contribution of entropy to the Helmholtz free energy for 
T-site and O-site can be estimated from the simple consideration that there are 
four times more configurations for T-sites. Therefore, T-sites are favored 
on the grounds of entropic arguments by $k_{B} T \log{4}$, 
approximately 34 meV at room temperature, which tends to make the
energetic more similar but still is not enough to make the T-site
the global minimum. 
Furthermore, zero-point energy corrections (ZPE) may affect differently
the energy of two absorption sites with different coordination and 
geometrical parameters.
A careful determination of the phonon spectra 
for this system demands the use of a considerable amount of computing time 
and is planned for a future paper. Our own initial estimates, based in
sketchy models that can be solved quickly, results in
the O-site getting further stabilized with respect to the
T-site by $\approx 20$ meV.
A word of caution is in order here since these values may have
a large error bar that we have not properly estimated. The model
used to deduce these values is as follows: the cut-off energy has
been reduced to $300$ eV, the k-points mesh to a mere 
$6 \times 6 \times 5$, phonons have been estimated only in the
$\Gamma$ point using a finite displacement method ($0.00529$ {\AA})
and a $2 \times 2 \times 2$ super-cell.
Therefore, these corrections are only to be taken as rough order 
of magnitude estimates, but in our calculations ZPE does not 
switch the preferred absorption site.
Frequencies in $\Gamma$ for the O and T sites are all real,
marking the character of global and local minima for these sites.

From a physical point of view, in this high concentration regime, 
the main characteristic of the tetragonal distortion necessary to 
accommodate H in the O-site is that it is both large and anisotropic. 
First, we notice that only occupation of similarly oriented O-sites over 
many unit cells would allow this kind of distortion.
Second, fractional changes in the UC volume
are large, $\frac{\Delta V}{V_{0}} =0.12$,
and $0.16$, for O-site and T-site respectively; this is
the trace of the strain tensor and measures the global stress
produced by the interstitial.
Therefore, it is necessary to understand the
stress distribution created around the distorted cell.
Total energies for a model where the UC is allowed
to relax keeping the cubic UC ($a=b=c$) show that when H is in O 
the system gains a fair amount of energy from the anisotropic deformation 
($510$ meV), while when H is near T-site the gain is smaller ($9$ meV). 
Under this cubic restriction, 
the fractional change of the UC volume
are $0.19$ and $0.16$ for O-site and T-site respectively; the interstitial
fits poorly in O in absence of the tetragonal distortion 
(internal pressure upon injection of the interstitial 
H in the clean iron BCC lattice amounts to several tens of GPa). 
From a chemical point of view, the role of the interstitial 
atom is to debilitate the Fe-Fe nearest-neighbors (NN) interaction; 
Fe-Fe bond length increases by $0.15$ and $0.08$ {\AA}
for H in O-site and T-site respectively. In both sites,
hydrogen tends to get negatively charged ($0.3 e$), while its spin is
delocalized contributing little to the total integrated
spin density that only increases by $\approx 0.15$ $\mu_{B}$ 
with respecto to clean BCC Fe. 
It is worth to notice that the tetragonal distortion of the
UC for H absorbed on the O-site ($\frac{c}{a} = 1.38$) 
is not far away from the distortion predicted for Bain's mechanism 
to transform from $\alpha$-iron to $\gamma$-iron
($\frac{c}{a}=1.41$)\cite{sliwko96}.

\begin{table}
\caption{
\label{TabGEOM1}
Optimum geometrical configurations for Fe$_{2}$H near
the octahedral and tetrahedral sites. Atomic positions,
$x$, $y$, $z$,
are given in fractional coordinates 
($\alpha=\beta=\gamma=90^{o}$). 
Total energies (in eV) correspond to the optimum geometry.
$F_{MAX}$ and $S_{MAX}$ are the maximum residual forces on
atoms and stresses on the cell. 
The charge transfer, $\Delta (q)$, is given in units of the electron
charge, and the spin is given in units of $\mu_{B}$ (Mulliken analysis).
}
\begin{tabular}{c}
 \hline
 \hline
 OCTAHEDRAL \\    
 \hline
   \begin{tabular}{cccc} \\
         energy (eV) & spin & F$_{MAX}$ (eV/{\AA}) & S$_{MAX}$ (GPa)  \\
         -1748.150    & 2.40 & $3 \times 10^{-6}$   & $4 \times 10^{-4}$  \\
   \end{tabular}    \\\hline
 \hline
   \begin{tabular}{cccccc} \\
  a ({\AA})  & b ({\AA})  & c ({\AA})  & c/a & V ({\AA}$^{3}$) &  $\Delta V/V$ (\%) \\
  2.628 & 2.628 & 3.618 & 1.38 & 24.99 & 0.12 \\
   \end{tabular}    \\\hline
 \hline
   \begin{tabular}{cccccc} \\
  atom      & x  & y  &  z & $\Delta (q)$ & spin \\
  H         & 0.5 & 0.5 & 0.0 & -0.30 & -0.02 \\
  Fe$_{a}$  & 0.0 & 0.0 & 0.0 & 0.19 & 1.11 \\
  Fe$_{b}$  & 0.5 & 0.5 & 0.5 & 0.10 & 1.29 \\
   \end{tabular}    \\\hline
 \hline
 \hline
 TETRAHEDRAL \\ 
 \hline
   \begin{tabular}{cccc} \\
         energy (eV) & spin & F$_{MAX}$ (eV/{\AA}) & S$_{MAX}$ (GPa)  \\
         -1748.097    & 2.46 & $1 \times 10^{-4}$   & $5 \times 10^{-4}$ \\
   \end{tabular}    \\\hline
 \hline
   \begin{tabular}{cccccc} \\
  a ({\AA})  & b ({\AA})  & c ({\AA})  & c/a & V ({\AA}$^{3}$) & $\Delta V/V$ (\%) \\
  2.896 & 2.984 & 2.984 & 0.97 & 25.78 & 0.16 \\
   \end{tabular}    \\\hline
 \hline
   \begin{tabular}{cccccc} \\
  atom      & x      & y  &  z & $\Delta (q)$ & spin \\
  H         & 0.2641 & 0.5 & 0.0 & -0.34 & -0.02 \\
  Fe$_{a}$  & 0.0    & 0.0 & 0.0 & 0.17 & 1.24 \\
  Fe$_{b}$  & 0.5283 & 0.5 & 0.5 & 0.17 & 1.24 \\
   \end{tabular}    \\\hline
 \hline
 \hline
\end{tabular}
\end{table}

\begin{figure}
\includegraphics[clip,width=0.80\columnwidth]{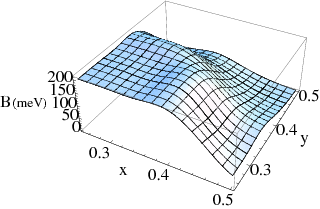}
\includegraphics[clip,width=0.80\columnwidth]{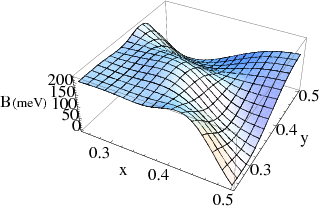}
\caption{
Total energy (meV) landscapes for (a) high
and (b) low concentrations cases (top and bottom panels). 
x and y are given in fractional units with respecto to the
$1 \times 1 \times 1$ BCC unit cell.
O-site is located at the top right corner 
$ \lbrace \frac{1}{2},\frac{1}{2},0 \rbrace$,
and T-site at the down right and top left corners.
$ \lbrace \frac{1}{4},\frac{1}{2},0 \rbrace$.
The region around the Fe at the origin (higher values
than $200$ meV) has been artificially put to a
constant value to improve visibility of the
relevant paths from O-site to T-site (a) and from T-site to T-site (b).
}
\label{FigLandScape}
\end{figure}
\begin{figure}
\includegraphics[clip,width=0.80\columnwidth]{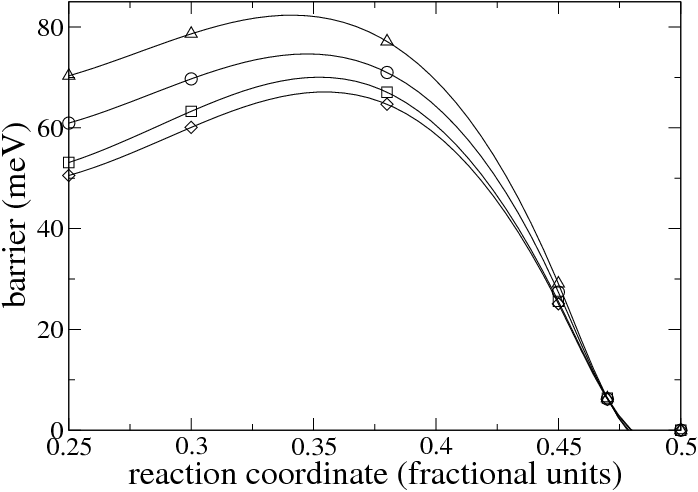}
\caption{
Diffusion barriers between T-site ($\frac{1}{4}$) 
and O-site ($\frac{1}{2}$) along the actual diffusion path
(see Fig.~\ref{FigLandScape}, the reaction coordinate has been scaled
between both endpoints)
for fully relaxed Fe$_{2}$H under different
external stresses. Triangles (a) and diamonds (d) correspond to hydrostatic 
pressures of $\pm 2$ GPa (tensile and compressive respectively), 
squares (c) to a tensile uniaxial stress of $-1$ GPa, and 
circles (b) to $0$ GPa.
The origin of energies has always been taken as the optimum equilibrium
value (for this concentration, the O-site). 
}
\label{FigBARRIERS}
\end{figure}

To compute diffusion barriers we have used
a linear synchronous-transit method followed by
quadratic refinement\cite{lst}. 
In addition, when a reduced number of parameters
could be established as the more relevant ones
exhaustive searches in regions of interest have been used
to display in more detail the topology around the
reaction pathway (Fig. ~\ref{FigLandScape}).
We have computed diffusion barriers between the
optimum configurations reached for H at O-site and T-site.
For the high density case, the transition state is 
located in the line joining O-site and T-site,
about half-way, with a barrier of $B=70$ meV 
(Fig.~\ref{FigBARRIERS}).  
Thermal diffusion can be estimated
by considering a typical vibrational frequency for
H in the lattice: $W \approx 10^{12}$ Hz.
Diffusion is quick,
hopping from O-site to T-site happens $D_{T}(B)=10^{11}$ times per second at 
room temperature. 
Quantum diffusion via tunneling through the barrier is important
for such a light atom as H. The tunneling rate can be estimated from a 
one-dimensional WKB approximation,
$
D_{Q}(B,d) = W e^{-4 d \sqrt{2mB}/3}
$,
where $d$ is the distance between minima ($0.6$ {\AA}). 
$D_{Q}$ is only lower than the thermal one, $D_{T}$, by a factor of 6.
At low temperature, the tunneling process becomes the
main mechanism, being equal in importance to the thermal one
at about $T \approx 183$ K, near the liquid Nitrogen temperature.
The effect of external stresses on the barrier is shown in 
Fig.~\ref{FigBARRIERS}.
As discussed above, ZPE can affect the energies of different
configurations, and in particular it can lower the already 
small diffusion barriers. 
The model described above, results in a lowering of the
barrier as seen from the O-site of $\approx 7$ meV, 
and a lowering of the barrier as seen from the T-site of $\approx 27$ meV.
We notice that frequencies computed for the TS now include a single imaginary
value, indicating that this is a first order saddle point.
Again, this correction should be taken as a crude estimate and 
we notice that similar calculations give a correction for the barrier 
of $46$ meV\cite{carter04}. 
We estimate that barriers corrected by ZPE
should be lower 
than the ones obtained from
total energies only 
by $\approx 40-50$\%.
A tensile stress of 2 GPa increases or decreases the
barrier by about 9\%; changing the thermal diffusion by about
30\% at room temperature, while the quantum mechanism is 
affected by 25\%. 
Barriers for this high concentration scenario 
have been computed using periodic boundary 
conditions on a $1 \times 1 \ times 1$ unit cell, representing 
diffusion of H and all the adjacent periodic copies simultaneously 
(notice that periodic boundary conditions respect the Fe$_{2}$H 
stoichiometry at all times during the diffusion process).
Finally, we check the internal consistency of these high-density simulations
by considering a $2 \times 2 \times 2$ unit cell, with the same Fe$_{2}$H
stoichiometry.
In this unit cell there are 36 O-sites and 144 T-sites that can be occupied 
by the 8 H. 
This makes $10^{7}$ possibilities for octahedral occupation, 
$10^{12}$ for tetrahedral occupation, and $10^{13}$ for 
an arbitrary mixture of both. 
A proper study of this system should try to sample these configurations, 
which is beyond the scope of this work. 
However, a check for the consistency of our previous results only needs to 
consider a few of these configurations. 
To ascertain the stability of the energy ordering between O and T-sites 
we consider two configurations where the octahedral/tetrahedral sites are 
occupied accordingly with results obtained in the $1 \times 1 \times 1$ unit 
cell (including the respective tetragonal distortions). 
The length of the vectors and all the atoms in the $2 \times 2 \times 2$ 
unit cell are allowed to relax 
in x, y and z directions including
72 degrees of freedom (the center of mass is constrained). 
A stationary point is considered converged when forces and stresses are below 
$0.002$ eV/{\AA} and $0.005$ GPa respectively. 
The O-site converges to a better minimum than the T-site by $0.470$ eV, 
which is consistent with the value derived from the $1 \times 1 \times 1$ 
simulation ($0.06 \times 8$) with an accuracy of 
1.2 meV/($1 \times 1 \times 1$ unit cell). 
In this model, if we try to compute the barrier for a single H hopping from 
the global minimum in the O-site to the T-site, 
we find that the effect of the occupation of the other 7 sites in octahedral 
positions is to transform this single T-site into a high order saddle point.
H in this T-site relaxes spontaneously to the neighboring O-site. 
Barriers computed with the help of the $1 \times 1 \times 1$ periodic model 
correspond to transforming large regions of all O-occupancy into large 
regions of all T-occupancy. 
We haven't investigated diffusion paths for H in the presence of a given 
mixture of O-sites and T-sites due to its combinatorial complexity.  
However, this example shows the interest to further study
$n \times n \times n$ cells in the high density regime in the future, 
and highlights the impact of the distortion of the lattice on the absorption 
energy on single local sites. 

\begin{table}
\caption{
\label{TabDFT}
For different concentrations and systems, energy differences
between the O-site and T-site sites ($\Delta E$ in meV), and fractional
changes in the unit cell volume for O-site and
T-site ($\frac{\Delta V}{V_{0}}$). 
}
\begin{tabular}{cc}
 \hline
DENSITY    
         & \begin{tabular}{rrrr} \\
         SYSTEM & $\Delta$E & $\frac{\Delta V}{V_{0}}\mid_{O}$ 
         $\frac{\Delta V}{V_{0}}\mid_{T}$ \\
        \end{tabular}    \\\hline
 Fe$_{2}$H 
         &  \begin{tabular}{rrrr} \\
         tetragonal & -53 & 0.12 & 0.16\\
         cubic & 452 & 0.19 & 0.16\\
         $1 \times 1 \times 1$ & 612 & 0.00 & 0.00 \\
        \end{tabular}    \\\hline
 Fe$_{16}$H 
         &  \begin{tabular}{rrrr} \\
         tetragonal & 119 & 0.03 & 0.03 \\
         cubic & 169 & 0.03 & 0.03 \\
         $2 \times 2 \times 2$ & 117 & 0.00 & 0.00 \\
        \end{tabular}    \\\hline
 Fe$_{54}$H 
         &  \begin{tabular}{rrrr} \\
         tetragonal & 105 & 0.004 & 0.003 \\
         cubic & 115  & 0.004 & 0.004 \\
         $3 \times 3 \times 3$ & 108 & 0.000 & 0.000 \\
        \end{tabular}    \\\hline
\end{tabular}
\end{table}

\section{Hydrogen in iron: low concentration and Finite Elements simulation}
We simulate the lower density scenario by placing one H in
the $2 \times 2 \times 2$ and 
$3 \times 3 \times 3$ unit cells  
(Fe$_{16}$H and Fe$_{54}$H).
At these concentrations, hydrogen is free to move around the
lattice without interacting with other hydrogen atoms (the remaining H-H
interaction on the $1 \times 1 \times 1$, $\approx 0.06$ eV, 
now becomes negligible). 
Other convergence parameters are kept the same, except for the k-points mesh, 
that has been reduced accordingly to the bigger size of the super-cell.
Now, the global total energy minimum still happens for a BCT lattice,
but it moves from the O-site to the T-site, that becomes
lower in energy w.r.t O by $119$ and $105$ meV for
Fe$_{16}$H and Fe$_{54}$H, respectively.
The change in the unit cell volume and the
tetragonal distortion of the lattice gets smaller with decreasing
density, and nearly independent of the absorption site,
$\frac{\Delta V}{V_{0}}= 0.03$ and $0.004$, and
$\frac{c}{a}=1.04$ and $1.003$ respectively
(Table~\ref{TabDFT}). 
The same calculation for a relaxed BCC lattice ($a=b=c=2.843$ {\AA})
yields a difference of $169$ meV in favor of T-site.
Diffusion happens now from T-site to T-site
with a barrier of $82$ meV (BCT) or $127$ meV (BCC)
(Fig.~\ref{FigLandScape}(b)).  
This barrier implies a diffusion rate $\approx 10$ 
times slower 
than for the high density case.
Tunneling is even more reduced, because the distance between
minima is larger ($\approx 0.9$ {\AA}).
In T-site, H moderately interacts with four NN at $1.65$ {\AA}, 
and gets negatively charged by $0.33 e$. 
In O, H interacts mainly with two NN, at $1.57$ {\AA}, while the 
nearest next-neighbors (NNN)
remain at $1.98$ {\AA} (H gets a similar amount of
charge, $0.32 e$).
Typical displacements for NN w.r.t. their positions in the clean
iron lattice are $\approx 0.1$ {\AA} (NN),
while NNN and more distant atoms move less than $\approx 0.02$ {\AA}. 
The spin on the different atoms does not show changes worthwhile to comment.


Strains to accommodate the interstitial H create a stress
distribution that has been analyzed by a finite elements (FE)
technique (input parameters have been obtained from
our calculations above). 
Assuming that deformations are elastic,
we have computed the stress distribution in the iron matrix
and the free energy variation
when H is absorbed in the O-site and T-site. 
Starting from strains obtained by DFT on the
$1 \times 1 \times 1$ super-cell, the FE simulation allows us
to study its effect in a surrounding region of 
size $n \times n \times n$
around the original volume ($n=2, 5$). 
Strictly speaking, this procedure implies introducing a deformation field
that has been computed 
using periodic boundary conditions on a bulk-like $1 \times 1 \times 1$ cell
into an infinite medium with the same elastic constants but
different boundary conditions (non-periodic).
This is therefore a slightly inconsistent procedure that
can only be justified 'a posteriori' by the agreement between the
energies obtained from the ab-initio calculations and the FE
approach (see below).

\begin{table}
\caption{
\label{TabFE}
Elastic energy distribution (meV), and fractional contribution
to the total one (\%) calculated by finite elements
over $n\times n\times n$  regions ($ n= 2, 3, 4, 5$).
Residual energy beyond $n=3$ is below a 3\% of the
total one. 
}
\begin{tabular}{ccccc}
\hline
 O-site 
   &  \begin{tabular}{ccccc} \\
       & $2\times $ & $3\times $ & $4\times $ & $5\times $ \\
   E( meV) & 195 & 215 & 219 & 221 \\
   \% & 88 & 9 & 2 & 0.8 \\
  \end{tabular}    \\\hline
 T-site 
   &  \begin{tabular}{ccccc} \\
       & $2\times $ & $3\times $ & $4\times $ & $5\times $ \\
   E( meV) & 23 & 24 & 24 & 24 \\
   \% & 95 & 4 & 0.7 & 0.3 \\
  \end{tabular}    \\\hline
\end{tabular}
\end{table}

\begin{figure}
\includegraphics[clip,width=0.80\columnwidth]{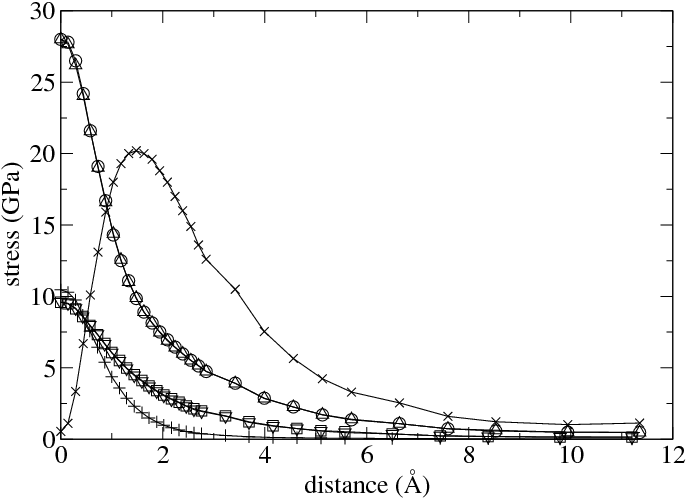}
\caption{
Finite elements simulation of stresses around the
fully relaxed $1 \times 1 \times 1$ unit cell upon absorption 
of H on the O-site (circles, $\circ$, and up-triangles,
$\bigtriangleup$, in the x and y
direction; times, $\times$, in the z one),
and T-site
(squares, 
$\Box$, 
and down-triangles, 
$\bigtriangledown$, in the x and y
direction; plus, $+$, in the z one).
}
\label{FigSTRESS}
\end{figure}

All the FE simulations have been performed with the program
Comsol Multiphysics\cite{finite}, using
$0.15$ {\AA} wide triangular elements.
Fig.~\ref{FigSTRESS} gives the diagonal components of the stress tensor 
for H in O-site and T-site
(the origin of distances is located at the surface of the deformed
$1 \times 1 \times 1$ region). Both cases show very little
residual stress beyond $6$ {\AA}  
(which corresponds to the $3 \times 3 \times 3$ supercell). 
The elastic energy distribution associated to these stresses
is well converged at these distances (Table~\ref{TabFE});
contributions from larger regions only amount to 1 or 2\% of
the total elastic energy. This result shows why it is not
necessary to go beyond the $3 \times 3 \times 3$ supercell in 
our ab-initio simulations. 
The stress distributions are very different for H in O-site and in T-site:
the former is very anisotropic
($x \sim y \neq z$)
and its maximum value
is $\approx 2$ times the maximum value for T-site.
The stored elastic energy, quadratic on the stress tensor, 
is approximately ten times higher for H in the O-site (Table~\ref{TabFE}).  
Therefore, storing one H atom in the O-site costs about $0.2$ eV 
to deform the surrounding cells if compared with
absorption in the T-site.
This number agrees well with the value previously obtained from the 
ab-initio simulation, $\approx 0.1$ eV for a fully relaxed lattice,
justifying the interpretation derived from these calculations. 
This approach shows that the main
difference between absorption in the T-site and O-site
is related to the relaxation of elastic energy in
neighboring cells. Only if all the cells are deformed
in the same way, i.e. when the system does not need to invest
energy again the elastic properties of the surrounding cells,
the O-site may become the preferred one. 

\section{conclusions}

For the case of high concentrations of interstitial H in BCC-iron,             
a large BCT distortion is found favorable with H absorbing
preferentially in the octahedral site.
This distortion is only possible if the hydrogen concentration
is high enough to transform a significant number of unit cells
from BCC to BCT. This transformation might form a precursor to
the BCC to FCC phase transformation.
For the case of low concentration of interstitial H, 
the energetic cost of deforming the surrounding
iron matrix makes preferable absorption in the T-site.
Hydrogen diffuses faster around regions of high concentration,
where the transition state is located between the T-site and O-site.
In low concentration regions, diffusion takes place
between T-site, avoiding the O-site (a local maximum).
H absorbed in O-site or T-site create a completely different distribution
of stresses in neighbouring cells explaining the preference
of H for the T-site for cases or regions of low concentration
of interstitials. 

\section{Acknowledgments}
This work has been financed by the Spanish
CICYT (MAT2003-3912 and MAT-2005-3866), 
and MEC (CONSOLIDERS SEDUREC and NANOSELECT). 
We are grateful with Dr. R. Ramirez for useful conversations.

\end{document}